\input{aipcheck}

\documentclass[
    ,final            
  ]
  {aipproc}

\layoutstyle{6x9}

\begin{document}

\title{Ab initio Stellar Astrophysics: Reliable Modeling of Cool White Dwarf Atmospheres
}


\author{Piotr M. Kowalski}{
  address={Helmholtz Centre Potsdam - GFZ German Research Centre for Geosciences, Telegrafenberg, 14473 Potsdam, Germany}
}

\author{Mukremin Kilic}{
  address={Smithsonian Astrophysical Observatory, 60 Garden Street, Cambridge, MA 02138, USA}
}

\begin{abstract}
Over the last decade {\it ab initio} modeling of material properties has become widespread
in diverse fields of research. It has proved to be a powerful tool 
for predicting various properties of matter under extreme conditions.
We apply modern computational chemistry and materials science methods, including density functional 
theory (DFT), to solve lingering problems in the modeling of the dense atmospheres of cool 
white dwarfs ($T_{\rm eff}\rm <7000 \, K$). Our work on the revision and improvements of the absorption mechanisms 
in the hydrogen and helium dominated atmospheres resulted in a new set of atmosphere models. By inclusion 
of the Ly-$\rm \alpha$ red wing opacity we successfully fitted the entire spectral energy distributions of known cool DA stars. 
In the subsequent work we fitted the majority of the coolest stars with hydrogen-rich models. This finding challenges
our understanding of the spectral evolution of cool white dwarfs. We discuss a few examples, 
including the cool companion to the pulsar PSR J0437-4715. 
The two problems important for the understanding of cool white dwarfs 
are the behavior of negative hydrogen ion and molecular carbon in a fluid-like, helium dominated medium. 
Using {\it ab initio} methods we investigate the stability and opacity of these two species in dense helium. 
Our investigation of $\rm C_2$ indicates that the absorption features observed in the ``peculiar''
DQp white dwarfs resemble the absorption of perturbed $\rm C_2$ in dense helium.
\end{abstract}

\maketitle


\section{Introduction}

\begin{figure}
\resizebox{\hsize}{!}{\rotatebox{270}{\includegraphics{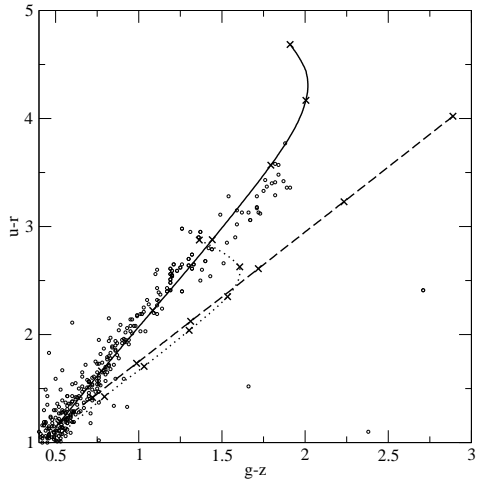}}}
\caption{The color-color diagram in the SDSS ugrz filters. The lines represent pure-hydrogen atmosphere 
models with (solid) and without (dotted) the Ly-$\alpha$ opacity and pure-helium atmosphere models (dashed). 
The crosses mark the values of the effective temperature ranging from $6000\rm \,K$ (bottom) 
to $3000\rm \,K$ (top) with decrements of $500\rm \, K$. The gravity is $\log g\rm=8\,(cgs)$. 
Circles are the data of \citet{Har06}. \label{F1}}
\end{figure}

Cool White dwarfs are among the oldest stars in the Galaxy. Being Gyrs old these stellar remnants are of importance 
for dating old stellar populations and understanding their star formation history. We are interested in understanding
the physical properties of the atmospheres and the spectra of cool white dwarfs. 
Over the last years we have introduced several improvements in the physics of dense media in our atmosphere models,
including the radiative transfer in the refractive atmosphere, the non-ideal chemical equilibrium abundances of spectroscopically important species and the opacity
of dense helium. Identification of the missing absorption mechanism at short wavelengths in the models of hydrogen-rich atmospheres 
as the Ly-$\alpha$ red wing opacity as well as better description of helium-rich atmospheres
have a significant impact on our understanding of cool white dwarfs. These improvements dramatically change the predicted flux at short wavelengths 
(especially in the Johnson U and B, and SDSS u and g filters). The hydrogen atmosphere models follow the observed sequence of cool white dwarfs on various color-color diagrams (Fig. \ref{F1}).
The best fits to the spectral energy distributions 
of the individual stars with $T_{\rm eff}\rm<6000\,K$ have been obtained with hydrogen-rich models. This suggest that most of the cool stars 
possess hydrogen-rich atmospheres, which challenges the current picture of the spectral evolution of cool white dwarfs.

In order to investigate the chemical evolution of white dwarf atmospheres the more extreme, fluid-like atmospheres of helium-rich stars have to be better understood. 
The atmospheres of these stars have extreme pressures and densities of a few $\rm g/cm^3$ and can not be described by ideal gas laws but must rather be treated as a helium-rich 
fluid in which the strong inter-particle interactions determine its physics and chemistry. The conditions prevailing at the surfaces of the helium-rich atmosphere stars 
could be diagnosed with trace species such as $\rm H^-$ and $\rm C_2$, through the impact of dense fluid on their absorption and the resulting white dwarf spectra.
Using modern methods we compute the properties of these species in dense helium. The resulted
high-pressure-induced spectral distortions are then matched with absorption signatures observed in the spectra of helium-rich stars 
(like distorted Swan bands in DQp stars) to derive the atmospheric parameters of the stars.

\section{Validation of the models}
\begin{figure}
\resizebox{\hsize}{!}{\rotatebox{270}{\includegraphics{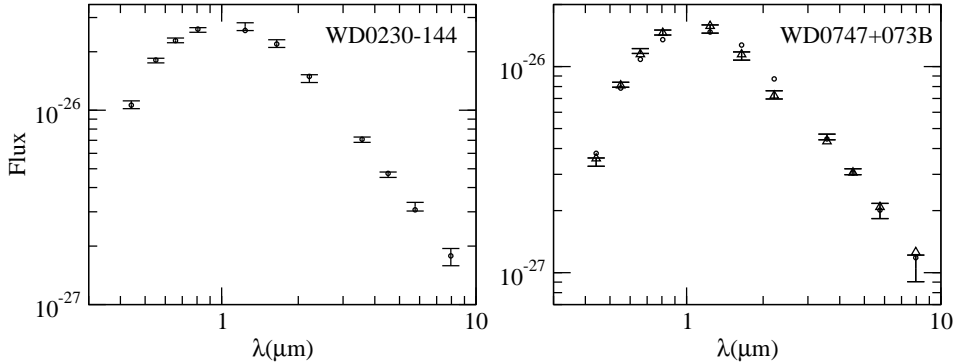}}}
  \caption{The fits to the SEDs of white dwarfs WD0230-144 and WD0747+073B. 
Open circles represent the fits by the pure-H models ($T_{\rm eff}{\rm =5474\,K,\,}\log g\rm=8.14$ and $T_{\rm eff}{\rm =4906\,K,\,}\log g\rm=8.12$ respectively),
while triangles represent the mixed He/H=0.83 solution for WD0747+073B 
($T_{\rm eff}{\rm =4697\,K,\,}\log g\rm=7.97$). The unit of the flux is $\rm erg \, cm^{-2}\,s^{-1}\,Hz^{-1}$. \label{F2}}
\end{figure}

With a grid of improved atmosphere models we have analyzed large samples of cool white dwarfs.
The colors of the new models are compared to a large SDSS sample in Figure \ref{F1}.
The new pure-H models not only reproduce the observed sequence of white dwarfs better than previous models but also show 
the need to include extreme line broadening of Ly-$\alpha$ to characterize white dwarfs with $T_{\rm eff}\rm<6000\,K$
when the short wavelength filters are used ($\lambda\rm<8000\,\mathring{A}$).
We also fitted the individual SEDs of a large sample of cool white dwarfs.
We found that most of the stars we fitted have pure-H atmospheres, 
with several having mixed H/He atmospheres dominated by H. Most of the stars
previously identified as having pure-He atmospheres \citep{Bergeron97,Bergeron01} turn out to be hydrogen-rich
stars in our analysis \citep{KS06,K08,K09a,K09b}.
Two examples of such fits are given in figure \ref{F2}. 
We also observe that the amount of helium in mixed stars seems to decrease with the decrease in the effective temperature
and the coolest stars ($T_{\rm eff}\rm <4500\,K$) possess pure-H atmospheres. 
The interpretation of these results requires further investigation and new, targeted observations 
to further validate the models.

\begin{figure}
\resizebox{\hsize}{!}{\rotatebox{270}{\includegraphics{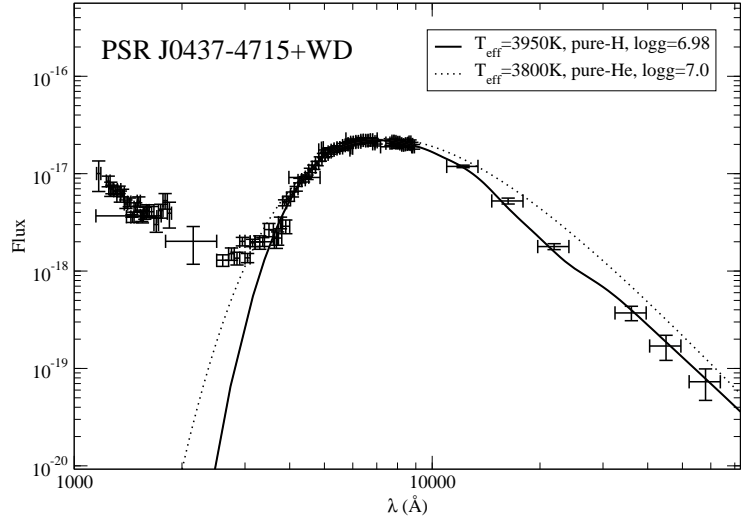}}}
  \caption{The measured spectrum of the pulsar and cool white dwarf binary system PSR J0437-4715 \citep{DK10}. Lines represent the fits by pure-H and pure-He models.
  The short wavelength flux is a contribution from the pulsar.
  The unit of the flux is $\rm erg \, cm^{-2}\,s^{-1}\,\mathring{A}^{-1}$. \label{F3}}
\end{figure}

The dependence of the derived atmospheric parameters on the atmosphere models used in the investigation is also evident for the low-mass, He-core white dwarfs.
These under-massive white dwarfs evolve through mass transfer in binary systems. 
One of the best known such system is a cool white dwarf orbiting the millisecond pulsar PSR J0437-4715. Using the atmosphere models of
\citet{BSW95} the white dwarf was assigned a helium-rich atmosphere \citep{B06}, which is at odds with the evolutionary models of such stars.
We fitted the entire spectral energy distribution of the star, including HST UV, optical,
near- and mid-infrared spectra with our models. The result is shown in figure \ref{F3}. An excellent fit is obtained with a pure-H atmosphere model
with $T_{\rm eff}\rm\sim4000\,K$. 
The presence of hydrogen is indicated by the flux suppression in the UV because the Ly-$\alpha$ opacity and near-IR because of CIA opacity from molecular hydrogen \citep{SJ99}, 
comparing with the fit by the pure-He atmosphere model.

\section{{\it Ab initio} investigations}
{\it Ab initio} methods represent a novel approach to the modeling of the properties of fluid-like
stellar atmospheres. In combination with the methods of the classical theory of fluids they are
a powerful tool for predicting the thermodynamical, optical and chemical properties of cool white dwarf atmospheres.
In our work we use the density functional theory, the most common {\it ab initio} method (PBE functional and plane wave codes, see \citet{DFT}).
Below we briefly discuss two examples of our recent {\it ab initio} work (see also Kowalski, these proceedings).

\subsection{$H^-$ in dense helium}
By the time it reaches $T_{\rm eff}\rm\sim6000\,K$ a white dwarf with an initial pure-He atmosphere should have accreted 
enough hydrogen from the interstellar medium for the latter to be detectable through the prominent
bound-free opacity from the negative hydrogen ion. \citet{Bergeron97} suggested that the negative 
hydrogen ion undergoes pressure ionization in the dense helium atmospheres, which could explain
the absence of hydrogen in the spectra of helium-rich, cool white dwarfs.
In order to address this problem, we performed quantum mechanical calculations of $\rm H^-$ in dense helium. 
Surprisingly, at least up to the density of helium of $4\rm \, g/cm^3$ (which is the highest density reached in the cool white dwarfs atmospheres) 
the negative hydrogen ion remains bound, which is indicated by the presence of two electrons around the proton.

\subsection{$C_2$ in dense helium}

\begin{figure}
\resizebox{\hsize}{!}{\rotatebox{270}{\includegraphics{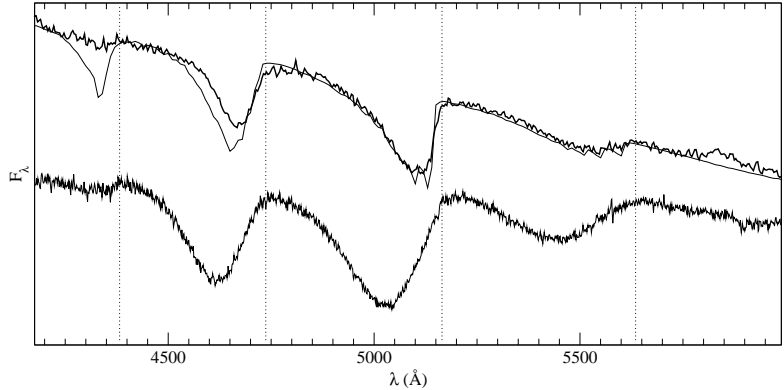}}}
  \caption{The optical spectra of the white dwarfs LHS179 (DQ, upper panel) and LHS290 (DQp, lower panel) \citep{Bergeron97}.
The solid line represents our fit to the spectrum of LHS179. The parameters of the fits are: $T_{\rm eff}\rm=6500 \, K$, log g=8 (cgs),
and $\log \, \rm C/He=-6.4$. 
The vertical dotted lines 
indicate the positions of the Swan bandheads for $\Delta\nu=\,+2,\,+1,\,0,\,-1$ from left to right respectively. 
\label{F4}}
\end{figure}

Some helium-rich atmosphere white dwarfs (DQ/DQp) show the $\rm C_2$ bands in the optical (Swan bands).
This gives the opportunity not only to investigate the properties of molecular carbon in their atmospheres, but to use $\rm C_2$ as a probe
of the physical conditions in cool helium-rich atmospheres, which are still poorly understood. There are two noteworthy observational facts 
about the DQ/DQp white dwarfs. Cutoff in their cooling sequence at $T_{\rm eff}\rm\sim6000\,K$ \citep{DB05,KK06} and the appearance of so called DQp stars, which show 
distorted Swan-bands-like spectral features in the optical, at lower effective temperatures \citep{SBF95,Bergeron97}. The examples of spectra of both types of stars are given 
in figure \ref{F4}. Two different explanations for the origin of the spectral features have been discussed in the literature.
\citet{LL84,HM08} speculated that the DQp stars show pressure shifted bands of $\rm C_2$, while other researchers
point to a different molecular species \cite{SBF95,Bergeron97,KB10}. The second idea was critically reviewed by \citet{HM08}.
To elucidate the nature of the spectral features observed in DQp stars we performed an extensive investigation of spectroscopic properties of $\rm C_2$ in dense helium, 
the electronic transition energy for the Swan bands and the vibrational frequencies, in dense fluid helium \citep{K10}. Our calculations show that the electronic transition energy for the Swan absorption 
increases significantly with the density ($\Delta T_e\rm\sim1.6\rho_{He}\,(eV)$), causing a blueward shift of the Swan bands in the spectra of DQp stars.
The observed shifts of about $0.08\rm\,eV$ for most of the DQpes indicate that 
the bands form at density of $\sim0.05\rm \, g/cm^3$, which is an order of magnitude lower 
than the one predicted by pure-He/C atmosphere models. This suggest pollution by hydrogen, which even in small amounts
increases the opacity and can lower the atmospheric density to the required values.

Our result shows that the optical bands observed in DQp stars are pressure shifted $\rm C_2$ bands
as was originally proposed by \citet{LL84}. The same approach to the $^3\Pi_u\rightarrow^3\Sigma^-_g$ Ballik-Ramsay absorption (10000-18000$\rm \, \mathring{A}$ bands, \citet{BR63}) gives
the blueward shifts of at least $\rm 1000 \, \mathring{A}$ ($\Delta T_{\rm e}\rm =2.2\rho_{He}\,(eV)$) for these bands in DQpes. The distortion of the Ballik-Ramsay bands
may produce the spectral depression observed at $\rm \sim 1\mu m$ for the DQp star LHS2293 \citep{F09}.

\section{Conclusions}
The atmospheres of cool white dwarfs could be understood only by having 
a reliable set of both hydrogen- and helium-rich atmosphere models, and modern {\it ab initio} methods 
of computational quantum mechanics are 
powerful tools to generate the necessary constitutive physics. 
With our improved models we have fitted the entire SEDs of various types of white dwarfs. 
Most of the white dwarfs with $T_{\rm eff}\rm <5000 \, K$ show H-rich atmospheric 
composition and the coolest stars could be fitted only with the pure-H models. This indicates that the accretion of hydrogen 
from the ISM plays an important role in the evolution of white dwarf atmospheres.
Our investigation of the properties of molecular carbon in the helium-rich atmosphere DQ/DQp stars shows that
the molecular bands observed in the optical spectra of DQp stars are the pressure shifted Swan bands.


\begin{theacknowledgments}
We thank Martin Durant, Oleg Kargaltsev, George Pavlov, Bettina Posselt, Marten van Kerkwijk and David Kaplan
for sharing the spectrum of PSR J0437-4715, Didier Saumon for useful comments on this paper and Sandro Jahn 
and Molecular modeling of geomaterials group at GFZ Potsdam for financial support.
\end{theacknowledgments}





\IfFileExists{\jobname.bbl}{}
 {\typeout{}
  \typeout{******************************************}
  \typeout{** Please run "bibtex \jobname" to optain}
  \typeout{** the bibliography and then re-run LaTeX}
  \typeout{** twice to fix the references!}
  \typeout{******************************************}
  \typeout{}
 }

\end{document}